\documentclass{iau} 
\usepackage{graphicx}
\usepackage[numbers]{natbib}
\usepackage{amsmath}

\usepackage{}

\usepackage[utf8]{inputenc}
\usepackage[swedish, english]{babel}
\usepackage[T1]{fontenc}

\usepackage{xcolor}

\title[SED model challenges at high $z$] %% give here short title %%
{Challenges in modelling the rest-frame ultraviolet/optical spectra of galaxies at the high-redshift frontier}

\author[Zackrisson \& Vikaeus]   %% give here short author list %%
{Erik Zackrisson$^1$  \and Anton Vikaeus$^1$}
\affiliation{$^1$Observational Astrophysics, Department of Physics and Astronomy,\\ Uppsala University, Box 516, SE-751 20 Uppsala, Sweden \\ email: {\tt erik.zackrisson@physics.uu.se}
}

\pubyear{2019}
\volume{341}  %% insert here IAU Symposium No.
\jname{Challenges in Panchromatic Galaxy Modelling with Next Generation Facilities}
\editors{M. Boquien, E. Lusso, C. Gruppioni, \& P. Tissera}
\begin{document}

\maketitle

\begin{abstract}
New challenges in the modelling of galaxy spectra are bound to emerge as upcoming telescopes like the James Webb Space Telescope will allow us to detect galaxies at fainter flux levels and higher redshifts than ever before. Here, we highlight three modelling problems that may become relevant for upcoming observations in the rest-frame ultraviolet/optical at the high-redshift frontier: stellar initial mass function sampling effects, Population III signatures and the leakage of ionizing radiation into the intergalactic medium.
\end{abstract}

\firstsection % if your document starts with a section,
\section{Introduction}
\noindent
Spectral synthesis models predict the spectra or spectral energy distributions (SEDs) of astronomical objects as a function of astrophysical parameters, either for the purpose of analyzing data, forecasting future observations or providing input for numerical simulations or other calculations. Current observations of the rest-frame ultraviolet(UV)/optical properties of galaxies at both low and high redshift have already revealed a number of challenges that need to be addressed in order for such codes to provide robust interpretations of galaxies at the high-redshift frontier. This includes problems in interpreting the rest-frame UV lines of the brightest Ly$\alpha$-emitters at z$\gtrsim 6$ \citep[e.g.][]{Stark16,Mainali18}, problems in explaining HeII emission properties at low metallicities \citep[e.g.][]{Senchyna17,Berg18}, and mismatches between the metallicities of stars and gas in intermediate-redshift galaxies \citep{Steidel16}. As new observational facilities like the \textit{James Webb Space Telescope (JWST)}, \textit{the Wide Field Infrared Survey Telescope (WFIRST)}, \textit{the Giant Magellan Telescope (GMT)}, \textit{the Thirty Meter Telescope (TMT)}, and \textit{the Extremely Large Telescope (ELT)} bring galaxies of unprecedented faintness at the high-redshift frontier within reach, one can also expect that the spectral synthesis community will soon have to face other modelling challenges. Here, we will highlight three such situations that we predict will become increasingly relevant for the high-redshift galaxy community in coming years: the modelling of Population III signatures, stellar initial mass function (IMF) sampling effects and Lyman continuum (LyC) leakage. Throughout this paper, we focus our discussion exclusively on galaxies at $z\gtrsim 6$, even though some of these issues are relevant at lower redshifts as well.

\section{Population III galaxies}
\noindent
The detection of signatures of chemically pristine stars, the so-far elusive Population III (hereafter Pop III), is sometimes referred to as the holy grail in the study of the high-redshifts Universe. Due to the cooling and fragmentation properties of zero-metallicity gas, the stellar initial mass function of Pop III stars is expected to be top-heavy, with a characteristic mass somewhere in the 1--1000 $M_\odot$ range \citep[e.g.][]{Stacy16,Hosokawa16}. 

While individual Pop III stars at high redshift may be beyond the reach of upcoming telescopes \citep[unless subject to extremely high magnification due to gravitational lensing;][]{Rydberg13,Windhorst18,Diego18}, there may be other viable avenues for studying the properties of such stars -- including Pop III supernovae (and their ejecta), Pop III GRBs -- and Pop III galaxies. The latter are small systems in which large numbers of Pop III stars manage to form within a limited time window, thereby potentially boosting the integrated luminosity of the Pop III stars to detectable levels. 

\begin{figure}
    \begin{center}
			\includegraphics[width=1.0\textwidth]{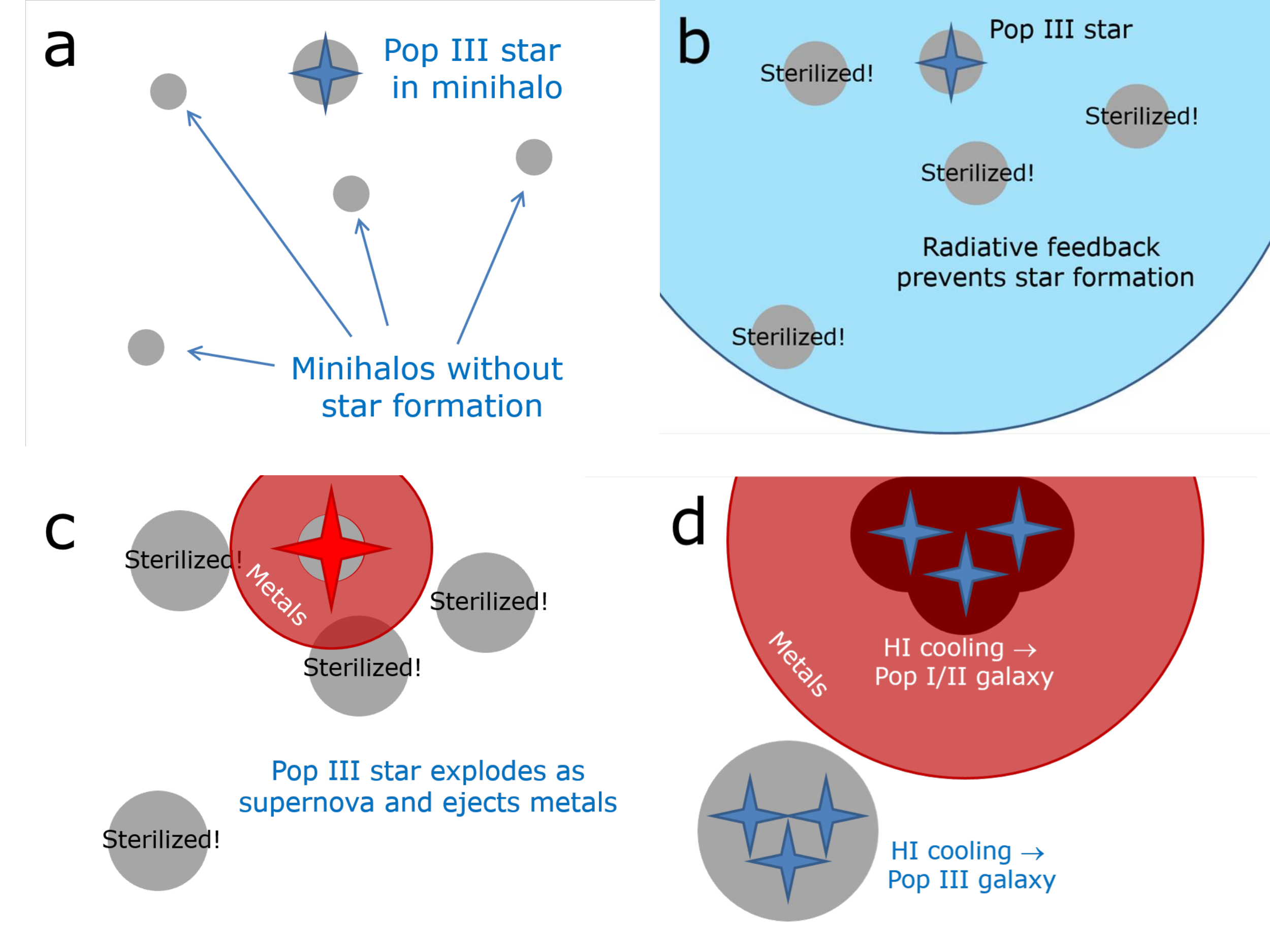}
		 \end{center}
    \caption{Schematic illustration of one potential scenario that leads to the formation of a Pop III galaxy. {\bf a)} A minihalo (a halo in the sub-HI cooling mass range) forms a single pop III star (or alternatively a small number of such stars), while adjacent minihalos at somewhat lower masses have yet to experience any star formation. {\bf b)} The Pop III star emits Lyman-Werner radiation (blue region) that inhibits star formation in the adjacent minihalos. {\bf c)} After having expired its lifetime, the pop III star goes supernova and ejects metals (red region) into the surrounding medium, while the nearby minihalos -- still incapable of forming stars -- continue to accrete material and increase in mass. {\bf d)} Several of the now metal-enriched minihalos merge with the post-Pop III minihalo and reach the HI cooling limit required for prolonged star formation. This object then forms a small galaxy, dominated by Pop I/II stars. A rare minihalo that has remained chemically pristine due to its distance from the Pop III supernova blast site, also reaches the HI cooling limit, but instead forms a Pop III galaxy.}
    \label{PopIII_formation}
\end{figure}

Pure Pop III galaxies may form from pristine gas in dark matter halos that have managed to reach the mass regime for HI cooling (total mass $\gtrsim 10^7$--$10^8\ M_\odot$, depending  on redshift) without experiencing prior star formation or being polluted by metals from nearby objects. Figure~\ref{PopIII_formation} features a schematic illustration of a possible scenario of this type. While such systems are expected to be quite numerous \citep{Stiavelli10}, most are expected to attain total stellar masses $\lesssim 10^4 M_\odot$ and hence lie well below the detection thresholds of currently planned telescopes (unless subject to very high magnifications along fortuitous sightlines; \citealt{Zackrisson15}). However, the stir in the community caused by the potential Pop III nature of the very bright Ly$\alpha$-emitter CR7 \citep{Sobral15} -- albeit no longer believed to be a Pop III galaxy \citep[e.g.][]{Sobral19} -- pushed the community to think hard about whether more massive systems could be created in rare situations. For instance, \citet{Visbal17,Yajima17,Inayoshi18} all argue that Pop III objects with total stellar mass $\sim 10^6\ M_\odot$ could potentially form. If the Pop III stars within one of these objects were to form within a time window of $\approx 10^7$ yr, with a stellar initial mass function centered around $\sim 10\ M_\odot$, the apparent rest-frame UV magnitude of such a galaxy would be $m_\mathrm{AB}\approx 30$--31 AB at $z\approx 6$--10. Due to the low probabilities of such relatively massive Pop III galaxies, large survey areas combined with a magnification boost due to strong lensing would likely be required to find them, and this may have to await the launch of WFIRST. However, \citet{Johnson18} have outlined a scenario in which star formation could be prevented in a metal-free halo close to a powerful quasar until the halo reaches a mass of $\sim 10^{10}\ M_\odot$. As such halos have the potential to form very massive Pop III galaxies (possibly with $m_\mathrm{AB}< 30$ mag), this motivates a search for such objects around known high-redshift quasars, which could be realized already with JWST.

\begin{figure}
    \begin{center}
			\includegraphics[width=0.9\textwidth]{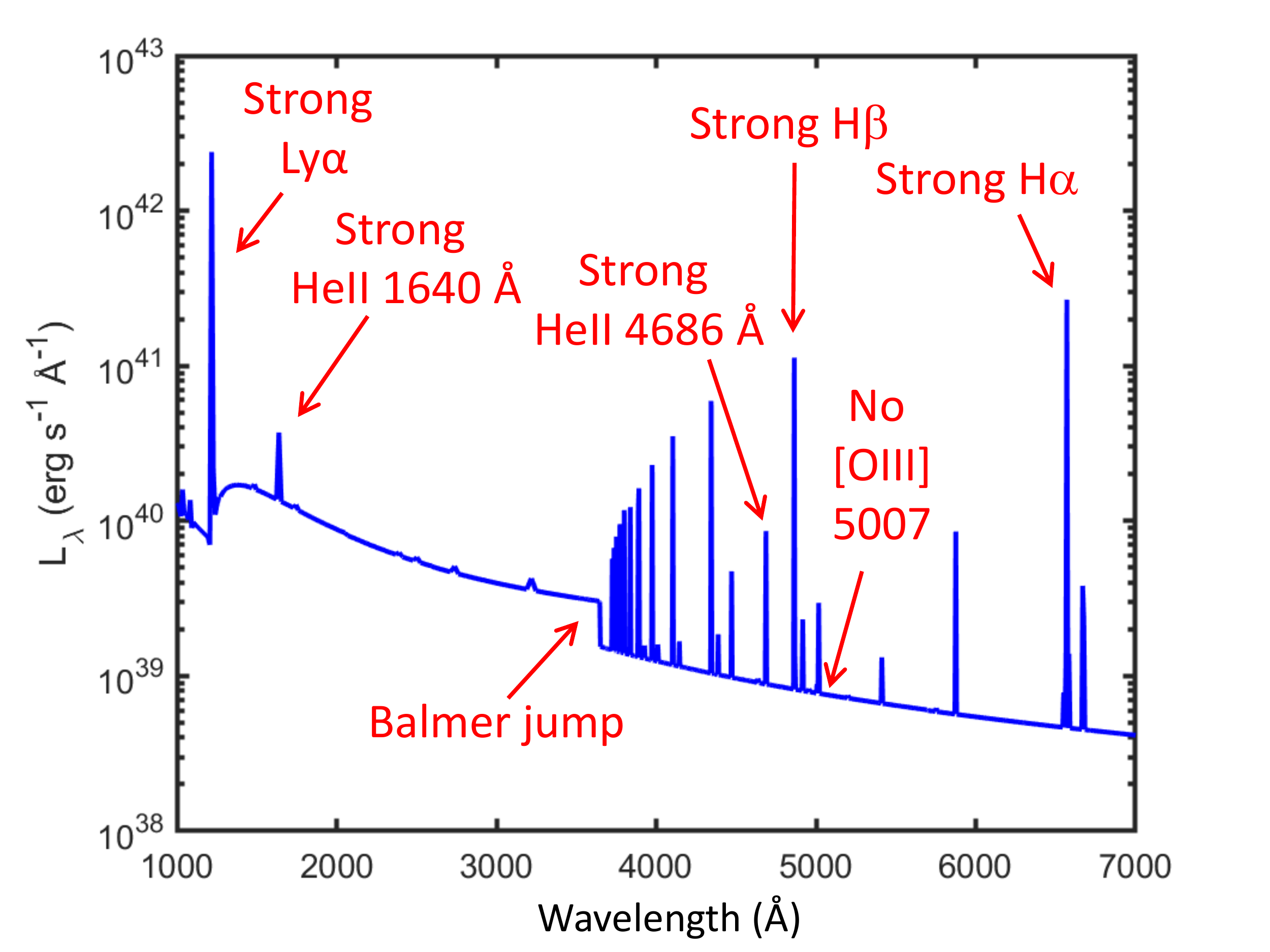}
    \end{center}
    \caption{Some of the rest-frame UV/optical Pop III spectral signatures discussed in the literature. Please note that Ly$\alpha$ (and the continuum at wavelengths directly shortward of this line) may be absorbed by the neutral intergalactic medium at $z>6$. This particular SED has been generated using the Yggdrasil \citep{Zackrisson11} model, under the assumption of a Pop III.1 stellar IMF (characteristic mass $\sim 100\ M_\odot$), an age of 1 Myr, a total stellar mass of $10^6\ M_\odot$ and negligible leakage of ionizing photons.}
    \label{PopIII_signatures}
\end{figure}

In coming years, new Pop III galaxy candidates are likely to be discovered, and developing SED models that can help determine the nature of such objects is an important task. Several spectral synthesis models geared to Pop III systems are already available, including those by \citet{Schaerer02,Schaerer03,Inoue11,Raiter10,Zackrisson11,Mas-Ribas16}, and several potential spectral diagnostics in the rest-frame UV/optical for such objects are known -- some of which are summarized in Figure~\ref{PopIII_signatures}. However, there are a number of predicted features of such systems have not yet made their way into the existing spectral synthesis models. 

Rotation creates increased mixing within Pop III stars and could, for sufficiently high rotation rates, lead to chemically homogenous evolution. In general, rotation acts to make Pop III stars live longer, attain higher luminosities and higher surface temperatures, which leads to boosted ionizing fluxes. As shown by \citet{Yoon12}, the hydrogen-ionizing fluxes may be boosted by factors of several for Pop III stars in the $\approx 20$--$300\ M_\odot$ range, provided that the rotational rate is sufficiently high, and the effects for He$^+$-ionizing radiation can be even more dramatic. Such effects have not yet been incorporated into spectral synthesis models for Pop III galaxies, although \citet{Schaerer03} does explore the consequences of a rotationally-induced Pop III Wolf Rayet phase. 

Because of recent interest in the potential gravitational wave signatures of merging black holes left behind from Pop III stars, a number of papers have attempted to model or simulate the Pop III binary population \citep[e.g][]{Kinugawa14,Belczynski17,Inayoshi17,Hirano18}. However, the impact that Pop III binary evolution may have on the spectra of Pop III galaxies have so far not been explored.

Some fraction of the ionizing radiation produced by Pop III stars could evade absorption in the dense interstellar medium in these objects, and instead contribute to the ionization of the intergalactic medium. While this is not expected to be a significant contributor to the reionization of the Universe (because Pop III galaxies are so rare), this would still alter the strength of emission lines and the ratio between direct star light and nebular emission in these objects. While some attempts to model a few different scenarios of this type have been made \citep{Inoue11,Zackrisson11}, the simulations by \citet{Johnson09} indicate that the escape could be highly selective, with hydrogen-ionizing radiation escaping more easily than He$^+$-ionizing radiation, with significantly altered emission-line ratios as a result. 

Once a candidate Pop III galaxy has been identified, possibly based on some combination of the spectral signature shown in Figure~\ref{PopIII_signatures}, there are primarily two questions that observers will attempt to answers: 1) Is this really a Pop III system; and 2) Is the stellar IMF top-heavy, as predicted by theory? While the solution to the first problem would be to establish that the system has a Pop III-like metallicity, detection thresholds for faint, high-redshift systems may make this far from trivial. Since Ly$\alpha$ may be absorbed in the neutral IGM at $z>6$, H$\alpha$ and [OIII]5007 are usually expected to be the strongest rest-frame UV/optical emission lines for reionization-epoch galaxies \citep{Shimizu16}. However, H$\alpha$ redshifts out of the reach of JWST/NIRSpec range at $z>7$, which may make H$\beta$ the strongest hydrogen recombination line that can be detected at the high-redshift frontier in the foresable future. If H$\beta$ is detected, but not [OIII], then this would be a strong indication of a very low metallicity, but how low? The exact value of the ratio [OIII]/H$\beta$ would be very hard to determine observationally for [OIII]/H$\beta$<0.1, which corresponds to [M/H]$\lesssim -3$ \citep{Inoue11}. However, this would still be consistent with a very metal-poor, yet definitely metal-enriched galaxy that has already started forming low-mass stars \citep[e.g.][]{Schneider12}. If no signatures of a top-heavy IMF (like a very strong HeII line \citep[e.g][]{Raiter10} is found in an object of this type, it would be hard to draw any conclusion on the Pop III IMF from this. 

Another problem lies in the fact that there may be a significant population of ``hybrid'' Pop III/II/I galaxies, i.e. galaxies that contains both metal-free and metal-enriched stars, as for instance predicted by the simulations of \citet{Sarmento18}. Another possibility is highlighted by the simulations by \citet{Xu16}, where a small halo forming Pop III stars photoionize gas that has been metal-enriched by other nearby halos. While the Pop III halo studied by \citet{Xu16} does not lie in the mass range where it would be readily detectable, one could envision situations like this giving rise to a Pop III galaxy that only contains Pop III stars (and, as a result, a very hard ionizing continuum) nonetheless displays metal emission lines. Problems of this type are already acute at low redshifts, for which there are claims that certain very metal-poor galaxies (but at a metallicity above the Pop III regime) display Pop III signatures within specific regions, primarily based on the strength of HeII emission lines \citep[e.g.][]{Kehrig15,Kehrig16,Ucci19,Kehrig18}. It is not immediately clear, however, that the discrepancy between the observed HeII fluxes and those predicted by models can be completely detached from the more general problem in explaining the strength of HeII emission in low-metallicity galaxies \citep[e.g.][]{Senchyna17,Berg18}.

\section{Stochastic IMF sampling}
\noindent
A standard assumption when modelling the SED of galaxies is that the stellar initial mass distribution can be treated as continuous, i.e. that the IMF is fully sampled at all points in time when new stars are formed. This assumption may break down in low-mass or low-SFR systems \citep[for a review, see][]{Cervino13}, with the result that two objects with identical ages, metallicities and star formation histories may exhibit different SEDs due to differences in the number of massive stars present at a given time. This problem is widely recognized in the study of local, low-mass systems like unresolved star clusters \citep[e.g.][]{Krumholz19}, but has been mostly ignored by the high-redshift community since the masses and SFRs of systems studied tend to be much too large for such effects to have any significant impact. However, as new telescopes push the detection thresholds to intrinsically fainter galaxies, IMF stochasticity may become an increasingly important problem in modelling and interpreting galaxy properties at high redshifts as well. \citet{Forero12} argue that this may have a noticeable effect on studies of Ly$\alpha$ equivalent widths and Lyman continuum leakage even for galaxies with SFRs as high as $\sim 1\ M_\odot$ yr$^{-1}$, and \citet{Paalvast17} predict non-negligible effects on the ratios of various strong optical emission lines for $< 0.1\ M_\odot$ yr$^{-1}$. It has also been argued that IMF sampling effects may also affect the impact of feedback recipes in simulations of low-mass galaxies (stellar mass $\lesssim 10^6\ M_\odot$) at $z\approx 0$ \citep{Su18} and galaxies close to the HI cooling limit ($\sim 10^8 \ M_\odot$) at $z\gtrsim 6$ \citep{Applebaum18}.

\begin{figure}
    \begin{center}
    \includegraphics[width=0.90\textwidth]{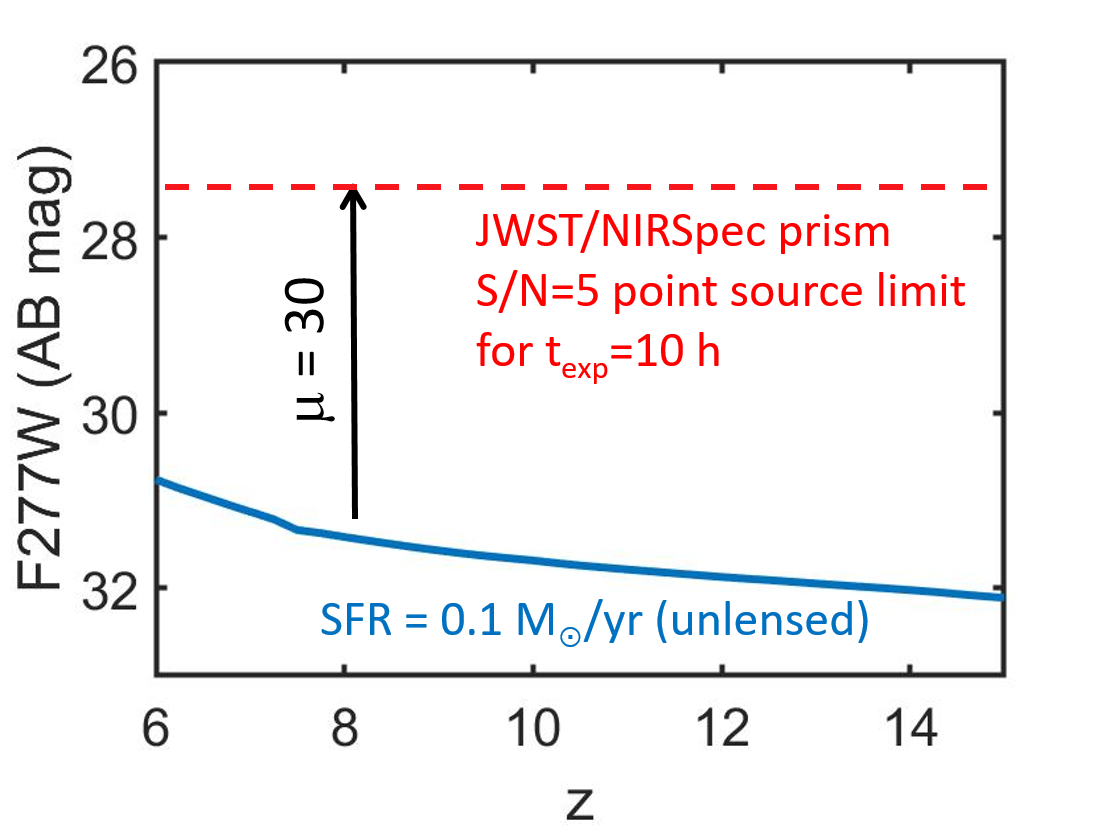}
    \end{center}
    \caption{The redshift dependence of the JWST/NIRCam F277W flux (measured in AB magnitudes) for a $Z=0.008$ galaxy that has experienced a constant SFR $= 0.1\ M_\odot$ yr$^{-1}$ for 100 Myr, compared to the JWST/NIRSpec spectroscopic $R=100$ point-source $S/N=5$ detection limit (dashed horizontal line) for an exposure time of 10 h. While the intrinsic flux of this galaxy (considered to be at the bright limit of the regime where IMF sampling effects may become important) lies comfortably below the JWST spectroscopic detection limit, strong lensing with magnification $\mu = 30$ (indicated by arrow) or more by a foreground galaxy cluster may well lift it into the detectable range. Hence, galaxies for which the SED is significantly affected by IMF sampling effects may turn up in deep spectroscopic studies of cluster fields with JWST.}
    \label{AB_limitfig}
\end{figure}

If we take SFR $\lesssim 0.1 \ M_\odot$ yr$^{-1}$ as the limit where IMF stochasticity may start to have non-negligible effects on galaxy SEDs \citep[e.g.][]{Paalvast17}, at what apparent magnitudes do we expect the affected systems to turn up at $z\gtrsim 6$? In Figure~\ref{AB_limitfig}, we plot the JWST/NIRCam F277W AB magnitudes for Yggdrasil \citep{Zackrisson11} $Z=0.008$ galaxies at $z=6$--15 that have experienced a constant SFR $= 0.1\ M_\odot$ yr$^{-1}$ for 100 Myr. As seen, the intrinsic AB magnitudes are in the range $m_{277}\approx 31$--32 mag\footnote{These galaxies are modelled assuming a fully sampled IMF, and IMF sampling effects would introduce slight galaxy-to-galaxy variations in their broadband fluxes.}, which -- in the absence of gravitational lensing -- is below the photometric detection threshold of even the longest JWST exposure times, and UV dust attenuation would make them even fainter still. However, such objects may still appear well above JWST detection thresholds in strongly lensed fields. For example, a magnification of $\mu>30$ could potentially push such systems above the JWST spectroscopic detection threshold ($m_\mathrm{AB}\approx 27.5$ mag for $S/N=5$ with JWST/NIRSpec prism after a 10 h exposure time). Hence, spectroscopic studies of $z\gtrsim 6$ galaxies in cluster lensing fields may well be probing galaxies for which sampling effects are significant.

\begin{figure}
    \begin{center}
    \includegraphics[width=0.9\textwidth]{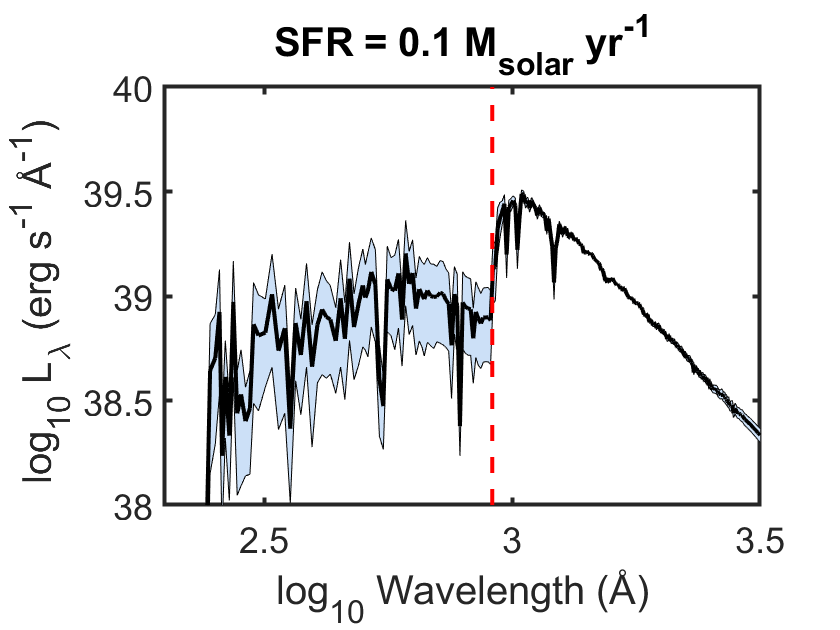}
    \end{center}
     \caption{The mean stellar population SED (black line), predicted by SLUG \citep{daSilva11} at rest-frame UV wavelengths for a $Z=0.15\ Z_\odot$ galaxy that has maintained a constant star formation rate at a level of SFR $= 0.1\ M_\odot$ yr$^{-1}$ for 100 Myr (giving a total stellar population mass of $10^7\ M_\odot$) in the case where are stars are born in stellar clusters. The coloured patch represents the galaxy-to-galaxy standard deviation of flux levels around the mean SED after a normalization to the average galaxy flux at 1500 \AA, indicating significant (factor of $\approx 2.5$) variations in ionizing flux (at wavelengths shortward of the Lyman limit at 912 \AA, marked by the vertical, red dashed line) despite potentially having similar 1500 \AA{} fluxes.}
    \label{SED_fig}
\end{figure}

What effect is this IMF stochasticity expected to have on such systems? In Figure~\ref{SED_fig}, we show the variations predicted in the stellar component of the rest-frame ultraviolet and optical spectra predicted by the SLUG code \citep{daSilva11} for an SFR $= 0.1\ M\odot$ yr$^{-1}$, $Z=0.15\ Z_\odot$ galaxy at an age of 100 Myr in the case where all stars are born in star clusters (thereby maximizing the impact of sampling effects). As seen, there are in this case highly significant (factor of $\approx 2.5$) galaxy-to-galaxy variations in the hydrogen-ionizing UV flux due to variations in the number of the most massive stars compared to the non-ionizing UV flux (which are produced by somewhat older stars of lower masses). This will result in variations in line equivalent widths (especially at wavelengths where the spectrum is not dominated by nebular continuum) that JWST will be able to study spectroscopically. Hence, the use of models that assume a stable star formation rate and a fully sampled IMF may lead to incorrect conclusions when applied to the analysis some of the faint, lensed galaxies that JWST will allow us to study in the years to come.

\section{Lyman-continuum leakage}
\label{LyC_leakage}
\noindent
If galaxies are to provide the ionizing photons for the reionization of the Universe, then some fraction of the LyC radiation produced by young, hot stars in these galaxies must be able to escape the interstellar medium and make it into the intergalactic medium. The LyC escape fraction $f_\mathrm{esc}$ required for cosmic reionization at $z>6$ is typically estimated at $f_\mathrm{esc}\approx 0.1$, but simulations often predict substantial temporal variations in this quantity, sometimes with escape fractions of $f_\mathrm{esc}>0.5$ attained for limited periods of time \citep[e.g.][]{Trebitsch17}. 

Since LyC leakage into the intergalactic medium (or the low-density circumgalactic medium) means that fewer ionizing photons will contribute to photoionization in the interstellar medium (where densities are high and recombination time scales short), LyC leakage lowers the relative contribution of nebular emission (in the form of emission lines and nebular free-bound/free-free continuum) to the observed galaxy SED. Depending on the escape mechanism, emission line ratios may also be affected. There are basically two different scenarios that can give rise to LyC escape \citep[e.g.][]{Zackrisson13}: an ionization-bounded nebula with holes (also known as the ``picket fence'' model) and a density-bounded nebula. To first order, the first mechanism will simply act to reduce all nebular contributions by a factor $1-f_\mathrm{esc}$. If the escape fraction is low ($f_\mathrm{esc} \lesssim 0.1$), this corresponds to a very small effect ($\lesssim 10\%$) on emission line fluxes and an even smaller effect on emission line equivalent widths (because part of the continuum may be nebular), which is often well within the observational error bars. Only at very high escape fractions ($f_\mathrm{esc}>0.5$) may the effects be sufficiently pronounced to give rise to tell-tale signatures. Density-bounded leakage, on the other hand, can give rise to detectable effects already at small $f_\mathrm{esc}$. For instance, \citet{Nakajima14} demonstrate how the [OIII]5007/[OII]3727 line ratio can change significantly even when going from $f_\mathrm{esc}=0$ to 0.1. In the context of classical HII regions, this happens because part of the [OII] zone of the nebula is truncated, whereas the [OIII] zone further in remains intact. 

While LyC leakage has so far not been considered an important ingredient in spectral modelling of galaxies, a failure to recognize cases of significant LyC leakage may lead to incorrect inference on star formation activity and nebular conditions for such objects. Accurately diagnosing LyC leakage is admittedly very challenging amidst the many other effects that can give rise to similar spectral signatures, especially given the small number of emission lines detectable at the high-redshift frontier. Changes in $f_\mathrm{esc}$ from ionization-bounded leakage with holes are for instance degenerate with rapid star formation fluctuations \citep[][]{Binggeli18}, which cause fluctuations in the ratio between the ionizing and the non-ionizing UV flux \citep[e.g.][]{Boquien14}, and hence in emission line strengths. The [OIII]/[OII] line ratios relevant for density-bounded leakage are also affected by the ionization parameter, the metallicity and the presence of shocks \citep[e.g][]{Allen08,Nakajima14}. Another problem is that some fraction of the ionizing photons may be absorbed by dust rather than gas, a scenario that can only be constrained through rest-frame far-infrared data \citep[e.g.][]{Inoue01,Hirashita03}, and even then only with great difficulty.

\end{document}